\documentclass{article}
\usepackage{amssymb}
\usepackage{amsfonts}
\usepackage{amsmath}
\usepackage{graphicx}

 \evensidemargin0in \oddsidemargin0in \topmargin10pt
\begin{document}

\title{Photon--added squeezed thermal states: statistical properties and its
decoherence in a photon-loss channel\thanks{{\small Work supported by the
National Natural Science Foundation of China (Nos.10775097 and 10874174).}}}
\author{Xue-xiang Xu$^{1,2}$, Li-yun Hu$^{1,2\thanks{
{\small Corresponding author. Tel./fax: +86 7918120370. Email address:
hlyun2008@gmail.com; hlyun2008@126.com. (L-Y Hu).}}}$, and Hong-yi Fan$^{1}$
\\
$^{1}${\small Department of Physics, Shanghai Jiao Tong University,
Shanghai, 200030, China;}\\
$^{2}${\small College of Physics and Communication Electronics, Jiangxi
Normal University, Nanchang, 330022, China.}}
\maketitle

\begin{abstract}
{\small Using the normally ordered Gaussian form of displaced-squeezed
thermal field characteristic of average photon number }$\bar{n}${\small , we
introduce the photon-added squeezed thermo state (PASTS) and investigate its
statistical properties, such as Mandel's Q-parameter, number distribution
(as a Legendre polynomial), the Wigner function. We then study its
decoherence in a photon-loss channel in term of the negativity of WF by
deriving the analytical expression of WF for PASTS. It is found that the WF
with single photon-added is always partial negative for the arbitrary values
of }$\bar{n}${\small \ and the squeezing parameter }$r${\small .}
\end{abstract}



\section{Introduction}

Nonclassicality of fields has been a topic of great interest in quantum
optics and quantum information processing \cite{1}. Experimentally, the
traditional quantum states, such as Fock states and coherent states as well
as squeezed states, have been generated but there are some limitations in
using them for various tasks of quantum information process \cite{2}.
Alternately, it is possible to generate and manipulate various nonclassical
optical fields by quantum superpositions and subtracting or adding photons
from/to traditional quantum states \cite{3,4,5,6,7,8,9,10,11,12,13,14,15,16}.

On the other hand, the single mode displaced, squeezed, mixed Gaussian
states have been paid enough attention by both experimentalists and
theoreticians. Marian et. al \cite{17,18} investegated the superposition of
a squeezed thermal radiation and a coherent one. They examined the squeezing
properties of the field using the distribution functions of the quadratures.
For Gaussian squeezed states of light, a scheme is also presented
experimentally to measure its squeezing, purity and entanglement \cite{19,20}%
. As is well known, dissipative quantum channels tend to deteriorate the
degree of nonclassicality (i.e., render quantum features unobservable).
Thus, it is usually necessary to investigate the decoherence properties in
dissipative channels, such as dynamical behaviors of the partial negativity
of Wigner function (WF) and how long a nonclassical field preserves its
partial negativity of WF. For instances, the nonclassicality of single
photon-added thermal states in the thermal channel is investigated by
exploring the volume of the negative part of the WF \cite{21}; Souza and
Nemes have derived an upper limit for the mixedness of the single bosonic
mode Gaussian states \cite{22} in a thermal channel.

In this paper, we shall introduce the photon-added squeezed thermo state
(PASTS) and investigate its statistical properties, such as Mandel's
Q-parameter, number distribution, the Wigner function. We then study its
decoherence in a photon-loss channel in term of the negativity of WF by
deriving the analytical expression of WF for PASTS. It is found that the WF
with single photon-added is always partial negative for the arbitrary values
of $\bar{n}$ and the squeezing parameter $r$. The work is arranged as
follows: In section 2, we introduce the state PASTS and derive its
normalized constant, and photon number distribution is discussed in section
3. Section 4 is devoted to calculating the WF. In the last section, we
explore the decoherence of PASTS in a photon-loss channel by discussing the
evolution of WF.

\section{PASTS and its normalization}

For a displaced-squeezed thermal field, the density operator is

\begin{equation}
\rho _{s}=D(\beta )S(r)\rho _{c}S^{\dagger }(r)D^{\dagger }(\beta ),
\label{1}
\end{equation}%
where $D(\beta )=\exp (\beta a^{\dagger }-\beta ^{\ast }a),$ and $S(r)=\exp
[ $i$r(QP+PQ)/2],$ are the displacement operator and the squeezing operator
\cite{23,24}, respectively, $\beta =(q+\mathtt{i}p)/\sqrt{2},$ $Q=\frac{%
a+a^{\dagger }}{\sqrt{2}},P=\frac{a-a^{\dagger }}{\sqrt{2}\mathtt{i}},\left[
a,a^{\dagger }\right] =1,$ and%
\begin{equation*}
\rho _{c}=(1-e^{-\frac{\hbar \omega }{kT}})e^{-\frac{\hbar \omega a^{\dagger
}a}{kT}},
\end{equation*}%
($k$ is the Boltzmann constant, $T$ denoting temperature), is qualified to
be a density operator of thermal (chaotic) field, since tr$\rho _{c}=1$. For
a coherent state $\left\vert z\right\rangle =\exp \left( za^{\dagger
}-z^{\ast }a\right) \left\vert 0\right\rangle $ \cite{25,26}, due to $%
a\left\vert z\right\rangle =z\left\vert z\right\rangle $, matrix elements of
any normally ordered operators $\colon \hat{O}\left( a^{\dagger },a\right)
\colon $ (the symbol $\colon \colon $denotes normally ordering) in the
coherent state is easily obtained, i.e,

\begin{align}
\left\langle z\right\vert \colon \hat{O}\left( a^{\dagger },a\right) \colon
\left\vert z^{\prime }\right\rangle & =O\left( z^{\ast },z^{\prime }\right)
\left\langle z\right. \left\vert z^{\prime }\right\rangle  \notag \\
& =O\left( z^{\ast },z^{\prime }\right) \exp \left\{ -\frac{\left\vert
z\right\vert ^{2}+\left\vert z^{\prime }\right\vert ^{2}}{2}+z^{\ast
}z^{\prime }\right\} ,  \label{2}
\end{align}%
so in Ref.\cite{27,28} by using the Weyl ordering invariance under
similarity transformations and the technique of integration within an
ordered product of operators (IWOP) Fan et al have converted $\rho _{s}$ to
its normally ordered Gaussian form

\begin{equation}
\rho _{s}=\frac{1}{\tau _{1}\tau _{2}}\colon \exp \left\{ -\frac{(q-Q)^{2}}{%
2\tau _{1}^{2}}-\frac{(p-P)^{2}}{2\tau _{2}^{2}}\right\} \colon ,  \label{3}
\end{equation}%
where
\begin{equation}
2\tau _{1}^{2}=(2\bar{n}+1)e^{2r}+1,2\tau _{2}^{2}=(2\bar{n}+1)e^{-2r}+1,
\label{4}
\end{equation}%
$\allowbreak $and $\bar{n}$ is the average photon number for $\rho _{c},$
i.e. $\bar{n}=(e^{\hbar \omega /kT}-1)^{-1}$ \cite{29}. The form in Eq.(\ref%
{3}) is similar to the bivariate normal distribution in statistics, which is
useful for us to further derive the marginal distributions of $\rho _{s}$.

Theoretically, the PASTS can be obtained by repeatedly operating the photon
creation operator $a^{\dagger }$ on a displacement squeezed thermal state,
so its density operator is defined as

\begin{equation}
\rho _{m}=N_{m}^{-1}a^{\dagger m}\rho _{s}a^{m},  \label{5}
\end{equation}%
where $m$ is a non-negative integer, $N_{m}=$tr$(a^{\dagger m}\rho
_{s}a^{m}) $ is the normalization constant. Using Eq.(\ref{3}) we known
immediately the normally ordered Gaussian form of $\rho _{m}$, i.e.,

\begin{equation}
\rho _{m}=\frac{N_{m}^{-1}}{\tau _{1}\tau _{2}}\colon a^{\dagger m}\exp
\left\{ -\frac{(q-Q)^{2}}{2\tau _{1}^{2}}-\frac{(p-P)^{2}}{2\tau _{2}^{2}}%
\right\} a^{m}\colon .  \label{6}
\end{equation}

Next we shall determine the normalization constant $N_{m}$. Using the
completness relation of coherent states $\int \frac{\text{d}^{2}z}{\pi }%
\left\vert z\right\rangle \left\langle z\right\vert =1$ as well as Eq.(\ref%
{2}), we have

\begin{eqnarray}
\mathtt{tr}\rho _{m} &=&\mathtt{tr}(\rho _{m}\int \frac{\text{d}^{2}z}{\pi }%
\left\vert z\right\rangle \left\langle z\right\vert )  \notag \\
&=&\frac{N_{m}^{-1}}{\tau _{1}\tau _{2}}\int \frac{\text{d}^{2}z}{\pi }%
\left\langle z\right\vert \colon a^{\dagger m}\exp \left\{ -\frac{(q-Q)^{2}}{%
2\tau _{1}^{2}}-\frac{(p-P)^{2}}{2\tau _{2}^{2}}\right\} a^{m}\colon
\left\vert z\right\rangle  \notag \\
&=&\frac{N_{m}^{-1}}{\tau _{1}\tau _{2}}\int \frac{\text{d}^{2}z}{\pi }%
z^{\ast m}z^{m}\exp \left[ -A\left\vert z\right\vert ^{2}+B^{\ast
}z+Bz^{\ast }+Cz^{2}+Cz^{\ast 2}+D\right] ,  \label{7}
\end{eqnarray}%
where we have set
\begin{eqnarray}
A &=&\frac{1}{2\tau _{1}^{2}}+\frac{1}{2\tau _{2}^{2}},B=\frac{1}{\sqrt{2}}%
\left( \frac{q}{\tau _{1}^{2}}+\frac{\mathtt{i}p}{\tau _{2}^{2}}\right) ,
\notag \\
C &=&-\frac{1}{4\tau _{1}^{2}}+\frac{1}{4\tau _{2}^{2}},D=-\frac{q^{2}}{%
2\tau _{1}^{2}}-\frac{p^{2}}{2\tau _{2}^{2}}.  \label{9}
\end{eqnarray}%
Due to $\mathtt{tr}\rho _{m}=1,$ thus we know
\begin{eqnarray}
N_{m} &=&\frac{e^{D}}{\tau _{1}\tau _{2}}\int \frac{\text{d}^{2}z}{\pi }%
z^{\ast m}z^{m}\exp \left\{ -A\left\vert z\right\vert ^{2}+B^{\ast
}z+Bz^{\ast }+Cz^{2}+Cz^{\ast 2}\right\}  \notag \\
&=&\frac{(-1)^{m}e^{D}}{\tau _{1}\tau _{2}}\frac{\partial ^{m}}{\partial
A^{m}}\int \frac{\text{d}^{2}z}{\pi }\exp \left\{ -A\left\vert z\right\vert
^{2}+B^{\ast }z+Bz^{\ast }+Cz^{2}+Cz^{\ast 2}\right\} .  \label{8}
\end{eqnarray}%
Further using the following integral formula \cite{30}
\begin{equation}
\int \frac{d^{2}z}{\pi }\exp \left\{ \zeta \left\vert z\right\vert ^{2}+\xi
z+\eta z^{\ast }+fz^{2}+gz^{\ast 2}\right\} =\frac{1}{\sqrt{\zeta ^{2}-4fg}}%
\exp \left\{ \frac{-\zeta \xi \eta +\xi ^{2}g+\eta ^{2}f}{\zeta ^{2}-4fg}%
\right\} ,  \label{11}
\end{equation}%
whose convergent condition is Re$\left( \zeta \pm f\pm g\right) <0$ and$\
\mathtt{Re}\left( \frac{\zeta ^{2}-4fg}{\zeta \pm f\pm g}\right) <0$, Eq.(%
\ref{8}) can be rewritten as follows
\begin{equation}
N_{m}=\frac{(-1)^{m}e^{D}}{\tau _{1}\tau _{2}}\frac{\partial ^{m}}{\partial
A^{m}}\left\{ \left( A^{2}-4C^{2}\right) ^{-1/2}\exp \left[ \frac{%
A\left\vert B\right\vert ^{2}+CB^{\ast 2}+CB^{2}}{A^{2}-4C^{2}}\right]
\right\} ,  \label{13}
\end{equation}%
which is the normalization constant of PASTS for photon-added number $m$. In
particular, when $\beta =0$ leading to $B=D=0,$ Eq.(\ref{13}) reduces to the
following form,

\begin{equation}
N_{m}=\frac{(-1)^{m}}{\tau _{1}\tau _{2}}\frac{\partial ^{m}}{\partial A^{m}}%
\left( A^{2}-4C^{2}\right) ^{-1/2}.  \label{14}
\end{equation}%
Especially when $m=0,1,2$ then the normalization constants are given by $%
N_{0}=1$, $N_{1}=\frac{1}{2}\left( \tau _{1}^{2}+\tau _{2}^{2}\right) $ and $%
N_{2}=\allowbreak \frac{1}{4}\left( 3\tau _{1}^{4}+2\tau _{1}^{2}\tau
_{2}^{2}+3\tau _{2}^{4}\right) \allowbreak $, respectively.

To see clearly the photon statistical properties of the PASTS, we will
examine the Mandel's $Q$-parameter defined as%
\begin{equation}
Q_{M}=\frac{\left\langle a^{\dag 2}a^{2}\right\rangle }{\left\langle a^{\dag
}a\right\rangle }-\left\langle a^{\dag }a\right\rangle ,  \label{15}
\end{equation}%
which measures the deviation of the variance of the photon number
distribution of the field state under consideration from the Poissonian
distribution of the coherent state. If $Q_{M}=0$ we say the field has
Poissonian photon statistics while for $Q>0$($Q<0$) we say that the field
has super-(sub-) Poissonian photon statistics.From Eq.(\ref{13}) and $N_{m}=$%
tr$(a^{\dagger m}\rho _{s}a^{m})$, we can easily calculate $\left\langle
a^{\dag }a\right\rangle =\frac{N_{m+1}}{N_{m}}-1,$ and $\left\langle a^{\dag
2}a^{2}\right\rangle =\frac{N_{m+2}}{N_{m}}-\frac{4N_{m+1}}{N_{m}}+2,$thus
we obtain the $Q$-parameter of the PASTS

\begin{equation}
Q_{M}=\frac{N_{m+2}-4N_{m+1}+2N_{m}}{N_{m+1}-N_{m}}-\frac{N_{m+1}-N_{m}}{%
N_{m}}.  \label{15a}
\end{equation}

It is well known that the negativity of the $Q_{M}$-parameter refers to
sub-Possonian statistics of the state. But a state can be nonclassical even
though $Q_{M}$ is positive. This case is true for the present state. From
Fig.1, one can clearly see that for the cases of $m=0$ (Fig.1(a))$,$ $Q_{M}$%
\ is always positive; while for $m\neq 0$ (for instance $m=1$) and a given $%
\bar{n}$ value, $Q_{M}$ becomes positive only when the squeezing parameter $%
r $ is more than a certain threshold value that increases as $m$ increases.
In addition, from Fig.1(a) and Fig.1(b) one can see that the threshold value
of $r$ decreases as $\bar{n}$ increases. We emphasize that the WF has
negative region for all $r$ and $\bar{n}$ thus the PASTS is nonclassical
(see next section below).

\section{Photon number distribution of PASTS}

In this section, we study photon number distribution of PASTS optical field.
Using the un-normalized coherent state $\left\vert \alpha \right\rangle
=\exp [\alpha a^{\dag }]\left\vert 0\right\rangle $, leading to $\left\vert
n\right\rangle =\frac{1}{\sqrt{n!}}\frac{\mathtt{d}^{n}}{\mathtt{d}\alpha
^{n}}\left\vert \alpha \right\rangle \left\vert _{\alpha =0}\right. ,$ $%
\left( \left\langle \alpha \right. \left\vert \alpha ^{\prime }\right\rangle
=e^{\alpha ^{\prime }\alpha ^{\ast }}\right) $, it is easy to see that the
photon number distribution formula is given by

\begin{eqnarray}
\mathcal{P}(n) &=&\text{tr}\left( \rho _{m}\left\vert n\right\rangle
\left\langle n\right\vert \right) =\left\langle n\right\vert \rho
_{m}\left\vert n\right\rangle  \notag \\
&=&\frac{1}{\tau _{1}\tau _{2}}\frac{1}{n!}\frac{\text{d}^{2n}}{\text{d}%
\alpha ^{\ast n}\text{d}\alpha ^{\prime n}}\left\langle \alpha \right\vert
\rho _{m}\left\vert \alpha ^{\prime }\right\rangle \left\vert _{\alpha
=\alpha ^{\prime }=0}\right. .  \label{16}
\end{eqnarray}%
Employing the normal ordering form of $\rho _{m}$ in Eq.(\ref{2}), Eq.(\ref%
{16}) can be put into the following form

\begin{equation}
\mathcal{P}(n)=\frac{N_{m}^{-1}e^{D}}{n!\tau _{1}\tau _{2}}\frac{\text{d}%
^{2n}}{\mathtt{d}\alpha ^{\ast n}\text{d}\alpha ^{\prime n}}\left\{ \alpha
^{\ast m}\alpha ^{\prime m}\exp \left[ B^{\ast }\alpha ^{\prime }+B\alpha
^{\ast }+C\alpha ^{\prime 2}+C\alpha ^{\ast 2}+\left( 1-A\right) \alpha
^{\ast }\alpha ^{\prime }\right] \right\} \left\vert _{\alpha =\alpha
^{\prime }=0}\right. .  \label{19}
\end{equation}%
Further expanding the exponential term $\exp \left[ \left( 1-A\right) \alpha
^{\ast }\alpha ^{\prime }\right] $ as series and using the generating
function of single-variable Hermite polynomials,
\begin{equation}
H_{n}(x)=\frac{\partial ^{n}}{\partial t^{n}}\exp \left( 2xt-t^{2}\right)
\left\vert _{t=0}\right. ,
\end{equation}%
we can calculate the photon number distribution (PND) of PASTS, i.e.,
\begin{align}
\mathcal{P}(n)& =\frac{N_{m}^{-1}e^{D}}{n!\tau _{1}\tau _{2}}%
\sum_{l=0}^{\infty }\frac{\left( 1-A\right) ^{l}}{l!}\frac{\partial
^{m+l}\partial ^{m+l}}{\partial B^{m+l}\partial B^{\ast m+l}}  \notag \\
& \times \frac{\text{d}^{2n}}{\text{d}\alpha ^{\prime n}\text{d}\alpha
^{\ast n}}\exp \left\{ B^{\ast }\alpha ^{\prime }+B\alpha ^{\ast }+C\alpha
^{\prime 2}+C\alpha ^{\ast 2}\right\} \left\vert _{\alpha =\alpha ^{\prime
}=0}\right.  \notag \\
& =\frac{N_{m}^{-1}\left\vert C\right\vert ^{n}e^{D}}{n!\tau _{1}\tau _{2}}%
\sum_{l=0}^{\infty }\frac{(1-A)^{l}}{l!}\left\vert \frac{\partial ^{m+l}}{%
\partial B^{m+l}}H_{n}\left[ \mathtt{i}B/(2\sqrt{C})\right] \right\vert ^{2}.
\label{20}
\end{align}%
After making the scale transformation and noticing the recurrence relation $%
\frac{\mathtt{d}^{l}}{\mathtt{d}x^{l}}H_{n}(x)=\frac{2^{l}n!}{\left(
n-l\right) !}H_{n-l}(x)$, we can easily obtain

\begin{equation}
\mathcal{P}(n)=\frac{N_{m}^{-1}e^{D}}{\tau _{1}\tau _{2}}\sum_{l=0}^{n-m}%
\frac{n!(1-A)^{l}\left\vert C\right\vert ^{n-m-l}}{l!\left[ \left(
n-m-l\right) !\right] ^{2}}\left\vert H_{n-m-l}\left[ \mathtt{i}B/(2\sqrt{C})%
\right] \right\vert ^{2},  \label{23}
\end{equation}%
Especially when $\beta =0$, Eq.(\ref{23}) reduces to

\begin{align}
\mathcal{P}(n)& =\frac{(-1)^{m}\sum_{j=0}^{\left[ \left( n-m\right) /2\right]
}\frac{n!(1-A)^{\left( n-m-2j\right) }\left\vert C\right\vert ^{2j}}{\left(
n-m-2j\right) !\left( j!\right) ^{2}}}{\frac{\partial ^{m}}{\partial A^{m}}%
\left( A^{2}-4C^{2}\right) ^{-1/2}}  \notag \\
& =\frac{n!\sigma ^{n-m}P_{n-m}(\frac{1-A}{\sigma })}{\left( n-m\right) !%
\frac{\partial ^{m}}{\partial A^{m}}\left( A^{2}-4C^{2}\right) ^{-1/2}},
\label{24}
\end{align}%
where%
\begin{equation}
\sigma =\sqrt{\left( 1-A\right) ^{2}-4C^{2}}=\sqrt{(\tau _{1}^{2}-1)(\tau
_{2}^{2}-1)}/(\tau _{1}\tau _{2}),
\end{equation}%
and in the last step of (\ref{24}) we have used the new expression of
Legendre polynomials \cite{8}%
\begin{equation}
P_{m}(x)=x^{m}\sum_{l=0}^{[m/2]}\frac{m!(1-\frac{1}{x^{2}})^{l}}{%
2^{2l}(l!)^{2}\left( m-2l\right) !}.
\end{equation}%
Eq. (\ref{23}) or (\ref{24}) is just the the analytical expression of the
PND of PASTS. In\ particular, when $m=0,$ Eq.(\ref{24}) becomes (with $\beta
=0$)

\begin{equation}
\mathcal{P}(n)=\frac{\sigma ^{n}}{\tau _{1}\tau _{2}}P_{n}(\frac{1-A}{\sigma
}).  \label{e26}
\end{equation}%
Eq.(\ref{e26}) is just the PND of the squeezed thermo state which seems a
new result. The PNDs of PASTS for some given parameters ($\bar{n},r$) and $m$
are plotted in Fig.2. From Fig. 2 it is found that the PND is constrained by
$n\geqslant m$. By adding photons, we have been able to move the peak from
zero photons to nonzero photons (see Fig.2 (a)-(c)). The position of peak
depends on how many photons are created and how much the state is squeezed
initially. In addition, comparing Fig.2(b) and Fig.2(d) we see that, for a
given $m$, the \textquotedblleft tail" of PND becomes more \textquotedblleft
wide" with the increasing parameter $r$.

\section{Wigner function of PASTS}

The Wigner function (WF) \cite{31} was first introduced by Wigner in 1932 to
calculate quantum correction to a classic distribution function of a
quantum-mechanical system. It now becomes a very popular tool to study the
nonclassical properties of quantum states. It is well known that WFs are
quasiprobability distributions because it may be negative in phase space
\cite{32}. Nevertheless, the partial negativity of the WF is indeed a good
indication of the highly nonclassical character of the state. Thus, to study
the dynamical behaviors of the partial negativity of WF and understand that
a nonclassical field preserves its partial negativity, Wigner distribution
may be very desirable for experimentally quantifying the variation of
nonclassicality \cite{33}.

The presence of negativity of the WF for an optical field is a signature of
its nonclassicality. In this section, using the normally ordered form of
PASTS, we evaluate its WF. For a single-mode system, the WF in the coherent
state representation $\left\vert z\right\rangle $ is given by \cite{34}

\begin{equation}
W(\alpha ,\alpha ^{\ast })=\frac{e^{2\left\vert \alpha \right\vert ^{2}}}{%
\pi }\int \frac{d^{2}z}{\pi }\left\langle -z\right\vert \rho _{m}\left\vert
z\right\rangle e^{-2\left( z\alpha ^{\ast }-z^{\ast }\alpha \right) },
\label{25}
\end{equation}%
where $\alpha =\left( x+\mathtt{i}y\right) /\sqrt{2}$. Then substituting Eq.
(\ref{6}) into Eq. (\ref{25}) and using Eq. (\ref{2}), we derive the WF of
PASTS%
\begin{equation}
W(\alpha ,\alpha ^{\ast })=\frac{N_{m}^{-1}e^{2\left\vert \alpha \right\vert
^{2}+D}}{\pi \tau _{1}\tau _{2}}\frac{\partial ^{m}}{\partial F^{m}}\left\{
\left( F^{2}-4C^{2}\right) ^{-1/2}\exp \left[ \frac{-F\left\vert
E\right\vert ^{2}+E^{\ast 2}C+E^{2}C}{F^{2}-4C^{2}}\right] \right\} ,
\label{26}
\end{equation}%
where we have set
\begin{equation}
F=2-A,\text{ }E=B-2\alpha ,
\end{equation}%
and used Eq.(\ref{11}). Especially when $\beta =0$, Eq.(\ref{26}) reduces to
\begin{equation}
W(\alpha ,\alpha ^{\ast })=\frac{(-1)^{m}e^{2\left\vert \alpha \right\vert
^{2}}}{\pi }\frac{\frac{\partial ^{m}}{\partial F^{m}}\left\{ \left(
F^{2}-4C^{2}\right) ^{-1/2}\exp \left[ \frac{-4F\left\vert \alpha
\right\vert ^{2}+4\alpha ^{\ast 2}C+4\alpha ^{2}C}{F^{2}-4C^{2}}\right]
\right\} }{\frac{\partial ^{m}}{\partial A^{m}}\left( A^{2}-4C^{2}\right)
^{-1/2}},  \label{27}
\end{equation}%
and further when $m=0,1$ Eq.(\ref{27}) becomes to%
\begin{equation}
W_{m=0}(\alpha ,\alpha ^{\ast })=\frac{1}{\pi (2\bar{n}+1)}\exp \left( -%
\frac{e^{-2r}x^{2}+e^{2r}y^{2}}{2\bar{n}+1}\right) ,  \label{28}
\end{equation}%
which is the WF of the squeezed thermo state, and
\begin{equation}
W_{m=1}(\alpha ,\alpha ^{\ast })=\frac{\mathfrak{M}x^{2}+\mathfrak{N}%
y^{2}+\Upsilon }{\pi }\exp \left( -\frac{e^{-2r}x^{2}+e^{2r}y^{2}}{2\bar{n}+1%
}\right) ,  \label{29}
\end{equation}%
respectively, where we have set

\begin{align}
\mathfrak{M}& =\frac{1}{\left( 2\bar{n}+1\right) ^{3}}\frac{\left( 2\tau
_{1}^{2}e^{-2r}\right) ^{2}}{\tau _{1}^{2}+\tau _{2}^{2}},  \label{30a} \\
\mathfrak{N}& =\frac{1}{\left( 2\bar{n}+1\right) ^{3}}\frac{\left( 2\tau
_{2}^{2}e^{2r}\right) ^{2}}{\tau _{1}^{2}+\tau _{2}^{2}},  \label{30} \\
\Upsilon & =\frac{1}{\left( 2\bar{n}+1\right) ^{3}}\frac{\tau _{1}^{2}+\tau
_{2}^{2}-4\tau _{1}^{2}\tau _{2}^{2}}{\tau _{1}^{2}+\tau _{2}^{2}}.
\label{30b}
\end{align}%
Eq.(\ref{28}) just agrees with the result of Eq.(48) in Ref. \cite{27},
whose form is normal distribution.

From Eq.(\ref{27}) one can see that the WF of the PASTS is always real, as
expected. When the factor $\mathfrak{M}x^{2}+\mathfrak{N}y^{2}+\Upsilon <0$
in Eq.(\ref{29}), the WF of the PASTS with $m=1$ has its negative
distribution in phase space. Noticing $\mathfrak{M,N}$ always positive, this
indicates that the WF of the PASTS always has the negative values under the
condition $\Upsilon <0$ (i.e., $\left( \tau _{1}^{2}+\tau _{2}^{2}-4\tau
_{1}^{2}\tau _{2}^{2}<0\right) $) at the phase space center $q=p=0$. In
fact, by substituting Eqs. (\ref{4}), (\ref{30a})-(\ref{30b}) into $\Upsilon
<0$, we find that for the arbitrary values of $\bar{n}$ and $r$, the WF with
$m=1$ is always partial negative.

Using Eq.(\ref{27}), the WFs of the PASTS are dipicted in phase space for
several different values of $m,\bar{n}$\ and $r$\ in phase space. Fig. 3
exhibits the WFs of the PASTS in phase space with $m=1$\ for different $\bar{%
n}$, $r$. It is easy to see that the WFs of single-PASTS always have the
negative region. The minimum in the negative region becomes larger with the
increasing of $\bar{n}$ (see Fig3.(a) and (c)). In Fig. 4, we have presented
the WFs with $\bar{n}=0.5$, $r=0.3$\ for different $m,$ which indicates that
the peak (absolute) value of WF become smaller as the increasing parameter $%
m $. The partial negativity of WF indicates the nonclassical nature of the
PASTS field.

\section{Evolution of WF in a photon-loss channel}

When the PASTS evolves in the amplitude decay channel, the evolution of the
density matrix can be described by the following master equation in the
interaction picture \cite{35},
\begin{equation}
\frac{d\rho }{dt}=\kappa \left( 2a\rho a^{\dagger }-a^{\dagger }a\rho -\rho
a^{\dagger }a\right) ,  \label{32}
\end{equation}%
where $\kappa $ represents the rate of decay. By using the thermal field
dynamics theory and thermal entangled state representation, the time
evolution of WF at time $t$ to be given by the following form \cite{36},
i.e.,%
\begin{align}
W\left( \alpha ,t\right) & =\frac{2}{\mathcal{T}}\int \frac{d^{2}z}{\pi }e^{-%
\frac{2}{\mathcal{T}}\left\vert \alpha -ze^{-\kappa t}\right\vert
^{2}}W\left( z,0\right) ,  \label{33} \\
\mathcal{T}& =1-e^{-2\kappa t}.  \notag
\end{align}%
Eq.(\ref{33}) is just the evolution formula of WF of single mode quantum
state in photon-loss channel. By observing Eq.(\ref{33}), we see that when $%
t\rightarrow 0,\mathcal{T}\rightarrow 0,$ $\frac{2}{\pi \mathcal{T}}\exp (-%
\frac{2}{\mathcal{T}}\left\vert \alpha -ze^{-\kappa t}\right\vert
^{2})\rightarrow \delta \left( \alpha -z\right) \delta \left( \alpha ^{\ast
}-z^{\ast }\right) ,$ so $W\left( \alpha ,t\right) \rightarrow W\left(
\alpha ,0\right) $ as expected. Thus the WF at any time can be obtained by
performing the integration when the initial WF is known.

For simplicity, here we only discuss the special case $\beta =0$.
Substituting Eq.(\ref{27}) into Eq.(\ref{33}) and using Eq.(\ref{11}), we
derive the time evolution of WF for PASTS in photon-loss channel:

\begin{equation}
W\left( \alpha ,t\right) =\frac{2(-1)^{m}}{\pi \mathcal{T}}\frac{\frac{%
\partial ^{m}}{\partial F^{m}}\left( \sqrt{%
\mathbb{N}
}e^{%
\mathbb{R}
\left\vert \alpha \right\vert ^{2}+\Bbbk \alpha ^{\ast 2}+\Bbbk \alpha
^{2}}\right) }{\frac{\partial ^{m}}{\partial A^{m}}\left(
A^{2}-4C^{2}\right) ^{-1/2}},  \label{35}
\end{equation}%
where we have set%
\begin{align}
\mathbb{N}
& =\frac{F^{2}-4C^{2}}{\left[ 4F+\left( \frac{2e^{-2\kappa t}}{\mathcal{T}}%
-2\right) \left( F^{2}-4C^{2}\right) \right] ^{2}-64C^{2}\allowbreak
\allowbreak },  \notag \\
\mathbb{R}
& =\frac{4%
\mathbb{N}
e^{-2\kappa t}}{\mathcal{T}^{2}}\left[ 4F+\left( \frac{2e^{-2\kappa t}}{%
\mathcal{T}}-2\right) \left( F^{2}-4C^{2}\right) \right] -\frac{2}{\mathcal{T%
}},  \label{36} \\
\Bbbk & =\frac{16%
\mathbb{N}
Ce^{-2\kappa t}}{\mathcal{T}^{2}}.  \notag
\end{align}%
In particular, when $m=0$ Eq.(\ref{35}) becomes
\begin{equation}
W_{m=0}\left( \alpha ,t\right) =\frac{2\sqrt{%
\mathbb{N}
\left( A^{2}-4C^{2}\right) }}{\pi \mathcal{T}}e^{%
\mathbb{R}
\left\vert \alpha \right\vert ^{2}+\Bbbk \alpha ^{\ast 2}+\Bbbk \alpha ^{2}}
\end{equation}%
which is just the WF of the squeezed thermo state in photon-loss channel.
This result can also be checked by substituting Eq.(\ref{28}) into Eq.(\ref%
{33}).

When $\kappa t$ exceeds a threshold value, the WF has no chance to be
negative in the whole phase space. At long time $\kappa t\rightarrow \infty
, $leading to $%
\mathbb{N}
\rightarrow \frac{1}{\allowbreak 4\left( F-2\right) ^{2}\allowbreak -16C^{2}}%
,$ $%
\mathbb{R}
\rightarrow -2,$ $\Bbbk \rightarrow 0,$ the WF in Eq.(\ref{35}) becomes

\begin{equation}
W\left( \alpha ,\infty \right) =\frac{e^{-2\left\vert \alpha \right\vert
^{2}}}{\pi },  \label{37}
\end{equation}%
which corresponds to the Gaussian state. In Fig.5, the WFs of PASTS are
depicted in phase space with $m=1$ and $r=0.3$ for several different $\kappa
t$. It is easily seen that the negative region of WF gradually disappears as
$\kappa t$\ increases. This implies that the system state reduces to a
Gaussian state after a long time interaction in the channel. Thus the loss
of channel causes the absence of the partial negativity of the WF if the
decay time $\kappa t$ exceeds a threshold value.

In Figs. 6, we have also presented the time-evolution of WF for different $r$%
. One can see clearly that the partial negativity of WF decreases gradually
as $r$ increases. The squeezing effect in one of the quadratures can be seen
in Fig.6. In Eq.(\ref{35}), for the PASTS we have obtained the expression of
the time evolution of WF. In principle, by differentiating as shown in Eq.(%
\ref{35}) we can derive the WF of other PASTS ($m\geqslant 1$). But its form
is too complicated. However, we can draw the Wigner distributions of PASTS
for $m=2,3,5$ by numerical simulation, as shown in Fig.7 (a)-(c),
respectively, from which one can see that the absolute value of the negative
minimum of the WF decreases as $m$ increases.

\section{Conclusions}

Based on the normally ordered Gaussian form of displaced-squeezed thermal
field, we have introduced a kind state: the photon-added squeezed thermo
state (PASTS). Then we have investigated the statistical properties of PASTS
(such as Mandel's Q-parameter, number distribution, the Wigner function) and
its decoherence in photon-loss channel with dissipative coefficient $\kappa $
in term of the negativity of WF by deriving the analytical expression of WF
for PASTS. It is found that the photon number distribution is just a
Legendre polynomial and that the WF with single photon-added is always
partial negative for the arbitrary values of $\bar{n}$ and the squeezing
parameter $r$. The technique of integration within an ordered product of
operators brings convenience in our derivation.

Figures caption:

Fig.1 Mandel's $Q$-parameter of PASTS as a fuction of $r$ with $%
m=0,1,5,10,30 $ (from top to bottom) for (a) $\bar{n}=0.3;$ (b) $\bar{n}=1.$

Fig.2 Photon number distributions of PASTS with $\bar{n}=1$ for (a) $%
r=0.3,m=0;$ (b) $r=0.3,m=1;$(c) $r=0.3,m=5;$(d) $r=0.8,m=1.$

Fig. 3 WF of PASTS for $m=1$ (a) $\bar{n}=0,r=0.3;$ (b) $\bar{n}=0,r=0.8;$%
(c) $\bar{n}=0.5,r=0.3;$(d) $\bar{n}=0.5,r=0.8.$

Fig. 4 WF of PASTS for $\bar{n}=0.5,r=0.3$ (a)$m=0;$ (b) $m=2;$(c)$m=3$ $;$%
(d) $m=5.$

Fig. 5 The time evolution of WF of PASTS for $m=1,r=0.3,$ and $\bar{n}=0.5$
with (a) $\kappa t=0.05;$ (b) $\kappa t=0.15;$(c) $\kappa t=0.2;$(d) $\kappa
t=0.4.$

Fig. 6 WF of PASTS for $m=1,$ $\bar{n}=0.5$ at $\kappa t=0.05,$ with (a) $%
r=0.01;$ (b) $r=0.5;$(c) $r=1.$

Fig. 7 WF of PASTS for $r=0.3$, $\bar{n}=0.5$ at $\kappa t=0.05,$ with (a)$%
m=2;$ (b) $m=3.$ (c) $m=5.$

\newpage

\begin{figure}[tbp]
\label{Fig1} \centering\includegraphics[width=12cm]{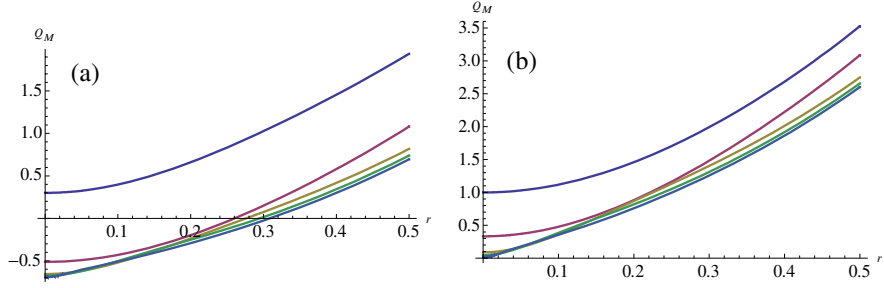}
\caption{Mandel's $Q$-parameter of PASTS as a fuction of $r$ with $%
m=0,1,5,10,30$ (from top to bottom) for (a) $\bar{n}=0.3;$ (b) $\bar{n}=1.$}
\end{figure}

\begin{figure}[tbp]
\label{Fig2} \centering\includegraphics[width=12cm]{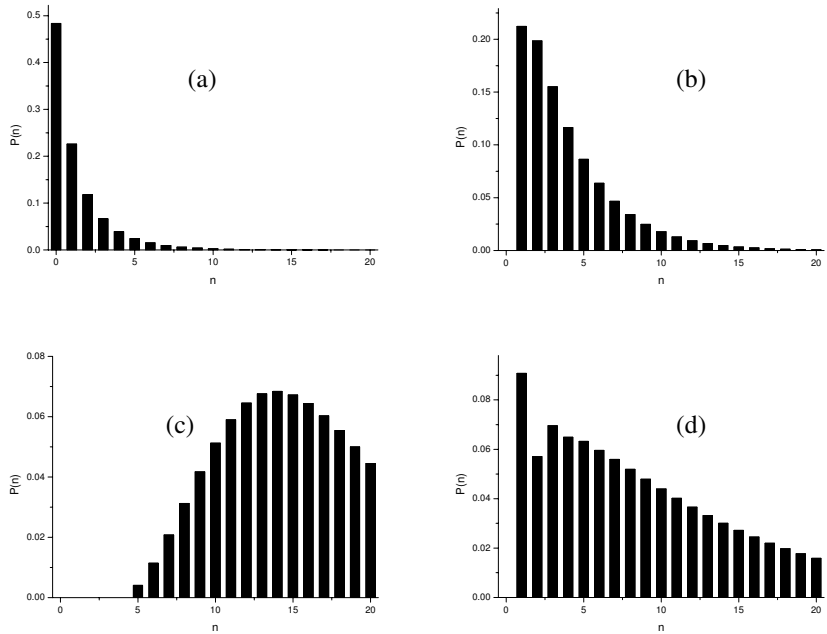}
\caption{Photon number distributions of PASTS with $\bar{n}=1$ for (a) $%
r=0.3,m=0;$ (b) $r=0.3,m=1;$(c) $r=0.3,m=5;$(d) $r=0.8,m=1.$}
\end{figure}

\begin{figure}[tbp]
\label{Fig3} \centering\includegraphics[width=12cm]{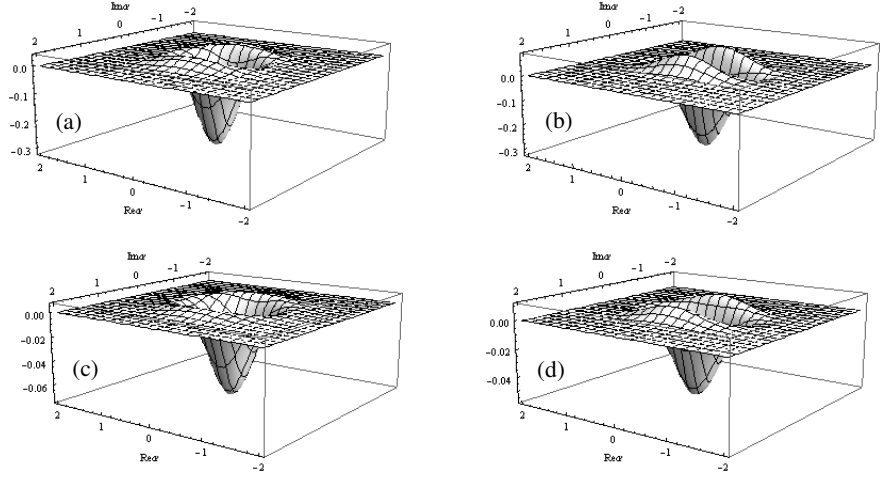}
\caption{WF of PASTS for $m=1$ (a) $\bar{n}=0,r=0.3;$ (b) $\bar{n}=0,r=0.8;$%
(c) $\bar{n}=0.5,r=0.3;$(d) $\bar{n}=0.5,r=0.8.$}
\end{figure}

\begin{figure}[tbp]
\label{Fig4} \centering\includegraphics[width=12cm]{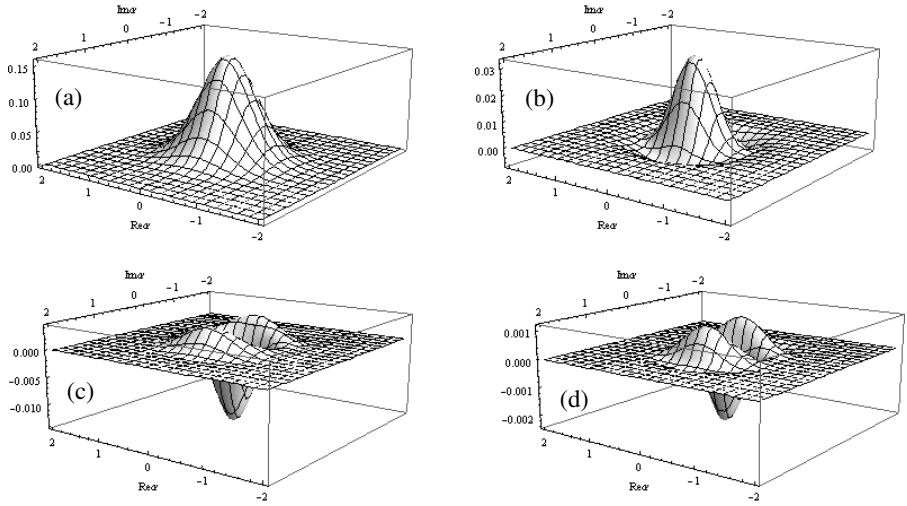}
\caption{WF of PASTS for $\bar{n}=0.5,r=0.3$ (a)$m=0;$ (b) $m=2;$(c)$m=3$ $;$%
(d) $m=5.$}
\end{figure}

\begin{figure}[tbp]
\label{Fig5}\centering\includegraphics[width=8cm]{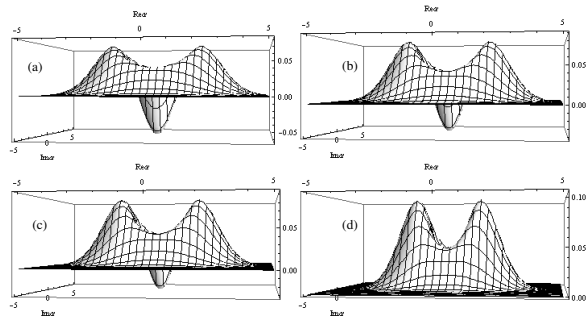}
\caption{The time evolution of WF of PASTS for $m=1,r=0.3,$ and $\bar{n}=0.5$
with (a) $\protect\kappa t=0.05;$ (b) $\protect\kappa t=0.15;$(c) $\protect%
\kappa t=0.2;$(d) $\protect\kappa t=0.4.$}
\end{figure}

\begin{figure}[tbp]
\label{Fig6}\centering\includegraphics[width=8cm]{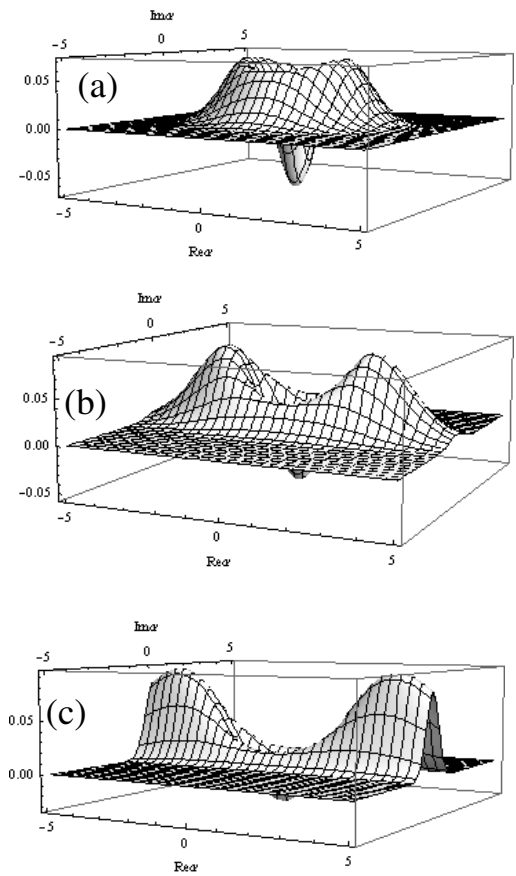}
\caption{WF of PASTS for $m=1,$ $\bar{n}=0.5$ at $\protect\kappa t=0.05,$
with (a) $r=0.01;$ (b) $r=0.5;$(c) $r=1.$}
\end{figure}

\begin{figure}[tbp]
\label{Fig7}\centering\includegraphics[width=8cm]{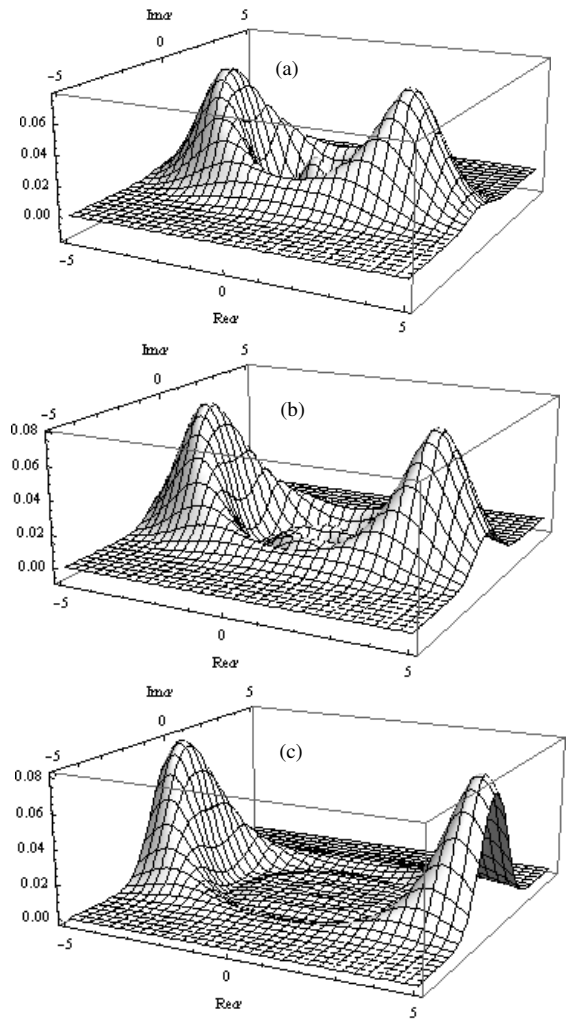}
\caption{WF of PASTS for $r=0.3$, $\bar{n}=0.5$ at $\protect\kappa t=0.05,$
with (a)$m=2;$ (b) $m=3.$ (c) $m=5.$}
\end{figure}

\end{document}